\documentclass{elsart}
\journal{Physics Letters A}

\usepackage{amssymb}
\usepackage{graphicx}

\newcommand{\ket}[1]{|\,{#1}\,\rangle}
\newcommand{\bra}[1]{\langle\,{#1}\,|}

\begin{document}

\begin{frontmatter}
\title{Nonlocality of a free atomic wave packet}

\author[ulm]{F. Haug\corauthref{cor}},
\corauth[cor]{Corresponding author.} 
\ead{florian.haug@physik.uni-ulm.de}
\author[ulm]{M. Freyberger}, 
\author[war]{and K. W\'odkiewicz}

\address[ulm]{Abteilung f\"ur Quantenphysik, Universit\"at Ulm, 
D-89069 Ulm, Germany} 
\address[war]{Instytut Fizyki Teoretycznej, Uniwersytet Warszawski, ul. Ho\.za 69, Warszawa 00-681, Poland}

\begin{abstract}
A simple model allows us to study the nonclassical behavior of slowly moving atoms interacting with a quantized field. Atom and field become entangled and their joint state can be identified as a mesoscopic ``Schr\"odinger--cat''. By introducing appropriate observables for atom and field and by analyzing correlations between them based on a Bell--type inequality we can show the corresponding nonclassical behavior.
\end{abstract}

\end{frontmatter}

\section{Introduction}

In 1935 Erwin Schr\"odinger published an influential paper \cite{schr:1935} whose contents is still a timely topic. In this work he introduced the term entanglement which lies at the very heart of quantum theory but has no counterpart in the classical description of nature. The measurement result found for one part of an entangled system depends in a nonclassical way on what is measured on the other part. This --- in a certain sense --- nonlocal feature of quantum mechanics has been pointed out first by Albert Einstein, Boris Podolsky and Nathan Rosen (EPR) in another famous paper of the year 1935 \cite{epr:1935}. In fact, Schr\"odinger himself called his paper a ``confession'' sparked by the EPR work. It was not until 1964 when John Bell succeeded to formulate the consequences of the EPR propositions on locality and reality \cite{bell:1964}. The limitations for specific correlations imposed by a classical theory based on locality and reality found expression in his seminal Bell--inequality. Their violation therefore provides a clear signature of nonclassical behavior of entangled systems \cite{werner:2001}.

The incompatibility between quantum mechanics and a classical description becomes also evident when one extrapolates the quantum mechanical description to macroscopic objects. Schr\"odinger highlighted this fact also in his paper of the year 1935 when he introduced his famous gedankenexperiment of an isolated cat in a box \cite{schr:1935}. The formulation of his gedankenexperiment results in the question whether superpositions like
\begin{equation}
\ket{\Psi} \equiv \frac{1}{\sqrt{2}} \left\{ \ket{alive}\ket{e} + \ket{dead} \ket{g}\right\}
\label{Eq:origcat}
\end{equation}
exist in nature or if they are just a mathematical consequence of quantum mechanics. Here, the states $\ket{alive}$ and $\ket{dead}$ mean macroscopic states of the ``Schr\"odinger cat'' correlated with the exited state $\ket e$ and the ground $\ket g$ of an atom \cite{schr:1935}. Since the work of Schr\"odinger it has become clear that such states can be at least prepared in the laboratory \cite{brun:1996,mon:1996,jul:2001,nak:1999,frie:2000,wal:2000} and possess the properties predicted by quantum theory. 

In this paper we present an atom--optic model which allows us to create at least mesoscopic Schr\"odinger--cat states. The cat--like degree of freedom will be replaced by the motion of an atom. Likewise we will point out the nonclassical aspect of this state by analyzing a Bell--type correlation sum. Hence it is the aim of the present paper to combine \cite{ban:1999,wod:2000a,wod:2000b} Schr\"odinger's cat with his term entanglement using only basic atom--optic elements \cite{ye:1999,pin:2000}. The used model is simple and only applies operationally defined elements.

The paper is organized as follows. In Sec.~\ref{sec:cat} we introduce a simple model consisting of a single two--level atom moving slowly through a quantized field. We show that the time evolution of the combined system leads to a state very similar to the original ``Schr\"odinger--cat'' state. In Sec.~\ref{sec:viol} we 
show how one can construct observables that demonstrate nonclassical correlations between the quantized atomic motion and the cavity field. Finally, in Sec.~\ref{sec:measure} we address the question if and in which way the observables constructed in the previous section have an operational meaning. We conclude with Sec.~\ref{sec:concl}.
\par 


\section{Atomic Schr\"odinger cat} 
\label{sec:cat} 

The aim of the following section is to briefly review a simple model based on atom optics which allows us to create states very similar to the original Schr\"odinger--cat state, Eq.~(\ref{Eq:origcat}).

\begin{figure}[ht]
\centerline{\includegraphics[width=9cm]{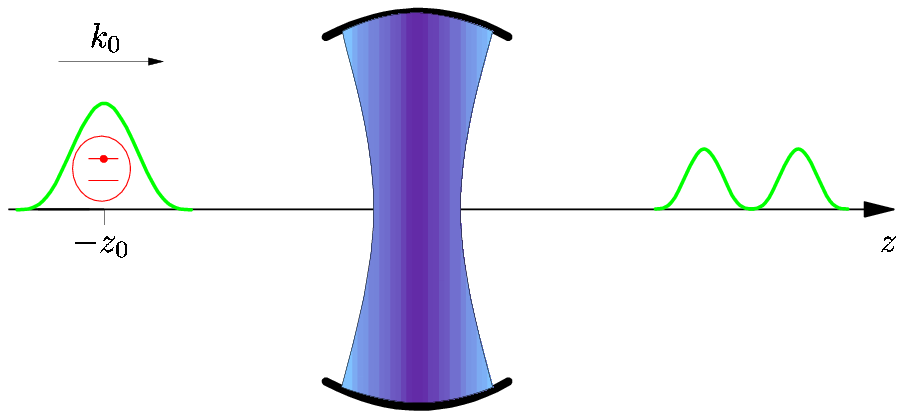}}
\caption{The Gaussian wave packet of a two--level atom in the excited state moves freely towards a cavity. The field inside the cavity is cooled down in the vacuum state. The atomic two--level structure and the quantized field create two effective optical potentials which split the wave packet in a superposition state.}
\label{fig:model}
\end{figure}

We consider \cite{stra:2000,stra:1999} a cold two--level Rydberg atom with ground state $\ket g$ and excited state $\ket e$ moving slowly in $z$--direction through a single--mode quantized field inside a cavity, as shown in Fig.~\ref{fig:model}. The atom interacts resonantly with the cavity field which is initially cooled down to the vacuum state $\ket 0$. We describe the initial center--of--mass motion of the atom by a Gaussian wave packet
\begin{equation}
\bra z\,\Phi\,\rangle = \Phi(z) = (2\pi\alpha)^{-1/4}\exp\left[-\frac{(z+z_0)^2}{4\alpha^2}+ik_0z\right]
\label{Eq:initgauss}
\end{equation} 
of width $\alpha$, centered at $-z_0$. At time $t=0$ the wave packet starts moving with mean velocity $\hbar k_0$ towards the cavity ``from the left to the right''.

Let us assume that atom and field are initially uncoupled.
When we start with an excited atom, the initial state of the complete system reads
\begin{equation}
\ket{\Psi(t=0)}=\ket\Phi\otimes\ket e\otimes\ket 0.
\label{Eq:initstate}
\end{equation}
In the interaction picture the time evolution of the combined system is described by the Hamiltonian
\begin{equation}
\hat H=\frac{\hat p_z^2}{2m} + \hbar\, \gamma\, u(\hat z)\left(\hat\sigma_-\hat a^\dagger + \hat\sigma_+\hat a\right).
\label{Eq:hamilt}
\end{equation} 
The atom of mass $m$ enters the cavity mode whose spatial structure is described by the mode function $u(z)$. The resonant interaction inside the cavity--field is modeled by a typical Jaynes--Cummings term \cite{schl:2001} with coupling strength $\hbar\gamma$. The operator--combination $\hat\sigma_-\hat a^\dagger$ describes the decay of an atom and the simultaneous creation of a cavity mode photon and has to be superposed with the combination $\hat\sigma_+\hat a$ standing for the excitation of the atom and the annihilation of a cavity mode photon. It turns out that exactly this resonant exchange of excitations between two--level atom and cavity field leads to a spatial splitting of the initial atomic wave packet $\ket\Phi$. Moreover, it leads to nonclassical entanglement between atom and field. 

To see this we look at the time evolution
$$
\ket{\Psi(t)} = \exp\left[-\frac{i}{\hbar}\hat{H}t\right]
\ket{\Phi}\ket{e}\ket{0}
$$
of our system in terms of the so--called dressed states 
$$
\ket{n,\pm} = \frac{1}{\sqrt{2}}\left\{\ket e \ket{n-1}\pm\ket g\ket n\right\},
$$
which combine the internal states of the atom with the Fock states $\ket{n}$ of the cavity field. Since the dressed states are the eigenstates of the Jaynes--Cummings part of the Hamiltonian, Eq.~(\ref{Eq:hamilt}), with eigenvalues $\pm \sqrt{n}$, we can appropriately expand the initial state $\ket{\Psi(t)}$, Eq.~(\ref{Eq:initstate}) and arrive at
\begin{equation} 
\ket{\Psi (t)} =
\frac{1}{\sqrt{2}}\Big[\ket{\Phi^{[+]}(t)}\ket{1,+} + \ket{\Phi^{[-]}(t)}\ket{1,-}\Big],
\label{Eq:dressedcat}
\end{equation}
where we have introduced the partial wave packets
$$
\ket{\Phi^{[\pm]}(t)} = \exp\left[-\frac{i}{\hbar}\left(\frac{\hat{p}_z^2}{2m} \pm
\hbar\gamma u(\hat{z})\right)t\right]\ket{\Phi}.
$$
Hence the complete time evolution of $\ket{\Psi(t)}$ can be reduced to the time evolution of the partial wave packets $\ket{\Phi^{[\pm]}(t)}$ in the effective optical potentials $\pm\hbar\gamma u(\hat z)$. Note that the atom ``sees'' both potentials at the same time. 

For a given mode function $u(z)$ the partial wave packets $\ket{\Phi^{[\pm]}(t)}$ can now be calculated. We concentrate here on the most simple case, that is we assume a square--well $u(z) = \Theta(L-|z|)$ of length $2L$ for the envelope of the cavity--field. It has been shown in \cite{stra:2000} that despite an additional phase the Gaussian form of the initial wave packet persists for both partial wave packets. On the right hand side of the cavity, that is for times $t\ge T$, where
$$
T \equiv \frac{2z_0}{\hbar k_0/m}
$$
represents the characteristic time needed by a classical particle moving form $-z_0$ to $+z_0$ with constant momentum $\hbar k_0$, the partial wave packets read
\begin{equation}
\bra z\,\Phi^{[\pm]}(t)\,\rangle = \Phi^{[\pm]}(z,t) = \frac{e^{i[\mp k_0D/2+\varphi(z,t;z_0\pm D/2)]}}{[\sqrt{2\pi}
(\alpha + i \frac{\hbar t}{2\alpha m})]^\frac{1}{2}}\exp\left[- \frac{(z+z_0-\frac
{\hbar k_0}{m}t\pm D/2)^2}{4\left(\alpha^2 + \frac{\hbar^2t^2}{4\alpha^2m^2}\right)}
\right].
\label{Gl:Phirechts}
\end{equation}
Here we have introduced the phase
$$
\varphi(z,t;z_0) \equiv k_0\left(z-\frac{\hbar k_0}{2m}t\right) + \frac{(z+z_0-
\frac{\hbar k_0}{m}t)^2}{4\alpha^2(\frac{2m\alpha^2}{\hbar t} + \frac{\hbar
t}{2m\alpha^2})}.
$$
The maxima of the two wave packets $\Phi^{[\pm]}(z,t)$, Eq.~(\ref{Gl:Phirechts}), are spatially well separated by the distance
\begin{equation}
D=\frac{\hbar\gamma}{\left(\frac{\hbar^2k_0^2}{2m}\right)}2L,
\label{Eq:distance}
\end{equation}
which scales with the length $2L$ of the interaction zone. Note that we have started initially with an empty cavity: even the vacuum state strongly influences the atomic motion and splits it into two spatially well separated parts \cite{har:1991,eng:1991}.

When we now compare the evolved state, Eq.~(\ref{Eq:dressedcat}), with the original Schr\"odinger--cat state, Eq.~(\ref{Eq:origcat}), we clearly recognize the same structure. In both cases we have a superposition of a mesoscopically or even macroscopically distinguishable state entangled with a microscopic state. The only difference is that in contrast to the original cat state, where cat and atom are two completely separate objects, here the dressed states involve both the internal atomic states as well as the states of the cavity field. In the following we will show the nonlocal and therefore nonclassical properties of this cat--like state by analyzing a Bell--type sum of correlation functions.
\par


\section{Nonlocal field--atom correlations}
\label{sec:viol}

The time evolution of our simple model--system presented in the previous section entangles all three degrees of freedom, i.e. the center--of--mass motion of the atom, its internal states and the cavity--field mode. Additionally, after passing the cavity the atom evolves freely in $z$--direction and is therefore in principle spatially well separated from the cavity. Nevertheless, measurement results found for the center--of--mass motion may depend nonlocally on what is measured on the internal states and the cavity--field mode. Such nonclassical correlations between subsystems can be shown using Bell--type inequalities \cite{bell:1964,werner:2001,chsh:1969}. 
Such inequalities always involve correlation functions of the form $\langle\hat A(\vec\alpha)\hat B(\vec\beta)\rangle\equiv\bra\Psi\hat A(\vec\alpha)\hat B(\vec\beta)\ket\Psi$ with hermitean and dichotomic operators $\hat A(\vec\alpha)$ and $\hat B(\vec\beta)$ describing measurements on a state $\ket\Psi$ depending on the measurement parameters $\vec\alpha$ and $\vec\beta$ respectively. In particular we will show that a CHSH--sum of correlations \cite{chsh:1969}
\begin{eqnarray}
{\mathcal B}_{\rm QM}(\vec\alpha,\vec\alpha',\vec\beta,\vec\beta') \equiv \langle \hat A(\vec\alpha)\hat B(\vec\beta)\rangle&+&\langle \hat A(\vec\alpha)\hat B(\vec\beta')\rangle \nonumber \\ +\langle \hat A(\vec\alpha')\hat B(\vec\beta)\rangle&-&\langle \hat A(\vec\alpha')\hat B(\vec\beta')\rangle
\label{Eq:qmCHSH}
\end{eqnarray}
can lead to a regime $|{\mathcal B}_{\rm QM}|> 2$ for our atom--field system and therefore clearly reveal its nonclassical behavior. For corresponding experiments on spin--like systems we direct attention to Refs.~\cite{asp:1982,wei:1998,tit:1998,row:2001} and references therein.
\par

\subsection{Observables and correlation function}

On the basis of the considerations of the previous section we are now in the position to show the nonlocal character of our model--system. For our purpose it is appropriate to rewrite the time--evolved state Eq.~(\ref{Eq:dressedcat}) in terms of the original internal and field states which yields 
\begin{eqnarray}
\ket{\Psi(t)}=\frac{1}{2}&\Big{[}&\left(\ket{\Phi^{[+]}(t)}+\ket{\Phi^{[-]}(t)}\right)\ket e\ket 0 \nonumber \\ &+& \left(\ket{\Phi^{[+]}(t)}-\ket{\Phi^{[-]}(t)}\right)\ket g\ket 1\Big{]}.
\label{Eq:violstate}
\end{eqnarray}
This state does not describe a discrete spin system but the continuous motion of an atom coupled to a quantized field. It has been pointed out earlier \cite{ban:1999,wod:2000a,wod:2000b,gil:1998,chen:2002} that continuous systems can violate Bell--type inequalities as well. Therefore, we face now the problem of finding proper dichotomic observables for the atomic motion, the internal atomic states and the cavity field which allow us to obtain nonclassical CHSH--correlations, Eq.~(\ref{Eq:qmCHSH}).

One of the simplest dichotomic operators for the atomic motion, i.e. for the partial wave--packets $\ket{\Phi^{[\pm]}}$ is the parity operator
\begin{equation}
\hat P\equiv\int\limits_{-\infty}^{\infty} dz\, \ket{-z}\bra{z}.
\label{Eq:parityop}
\end{equation}
As we know from Eq.~(\ref{Eq:qmCHSH}) we need adjustable parameters $\vec\alpha$, $\vec\beta$, etc. in order to define the reference frame of the measurement. It is probably most natural \cite{ban:1999,wod:2000a,wod:2000b} when we look at parity from different points $(z,k)$ in phase--space by displacing the parity with the unitary displacement operator
\begin{equation}
\hat D(z,k)\equiv \exp{\left[i(k\hat z-z\hat k)\right]},
\label{Eq:displop}
\end{equation}
where $\hat z$ is the position operator and $\hat k = \hat p/\hbar$ represents the scaled momentum operator. 

Although our idea is to show the nonclassical character of the CHSH--sum, Eq.~(\ref{Eq:qmCHSH}), with the help of continuous degrees of freedom, i.e with the atomic center--of--mass motion and the cavity field we have to deal with the internal states. A simple projection to the ground or the excited state does not come into question since it would clearly destroy the entanglement between atomic motion and cavity field. Instead we consider a superposition measurement described by the hermitean operator 
\begin{equation}
\hat\sigma\equiv\ket g\bra e + \ket e\bra g.
\label{Eq:intop}
\end{equation}

Merging the operators Eqs.~(\ref{Eq:parityop}), (\ref{Eq:displop}) and (\ref{Eq:intop}) we arrive finally at the observable 
\begin{equation}
\hat{\mathcal A}(z,k)\equiv\hat D(z,k)\hat P\hat D^\dagger(z,k)\otimes \hat\sigma
\label{Eq:atomop}
\end{equation}
for the atom.

Due to the chosen initial conditions, Eq.~(\ref{Eq:initstate}), the whole setup consisting of atom and field contains only a single excitation. Therefore, for the field we can construct as well a dichotomic hermitean operator of the form
\begin{equation}
\hat{\mathcal B}(\beta)\equiv e^{i\beta}\ket 0\bra 1 + e^{-i\beta}\ket 1\bra 0
\label{Eq:fieldop}
\end{equation} 
depending on the parameter $\beta$.

Having introduced all necessary ingredients we are now in the position to combine Eqs.~(\ref{Eq:violstate}), (\ref{Eq:atomop}) and (\ref{Eq:fieldop}) and calculate the correlation function
$$
\langle\hat{\mathcal A}(z,k)\hat{\mathcal B}(\beta)\rangle \equiv 
\bra{\Psi(t)}\hat{\mathcal A}(z,k)\hat{\mathcal B}(\beta)\ket{\Psi(t)},
$$
which depends on three real parameters: The pair $(z,k)$ parameterizes the reference frame of our first subsystem, namely the atom, while the phase $\beta$ parameterizes the measurements on the second subsystem which is the cavity field. 

We can further analyze this correlation function 
\begin{eqnarray}
\langle\hat{\mathcal A}(z,k)\hat{\mathcal B}(\beta)\rangle &=& \frac{1}{2}\left(\bra{\Phi^{[+]}}\hat {\mathcal W}(z,k) \ket{\Phi^{[+]}}-\bra{\Phi^{[-]}}\hat {\mathcal W}(z,k)\ket{\Phi^{[-]}}\right)\cos\beta \nonumber \\
&-&\frac{i}{2}\left(\bra{\Phi^{[+]}}\hat {\mathcal W}(z,k)\ket{\Phi^{[-]}}-\bra{\Phi^{[-]}}\hat {\mathcal W}(z,k) \ket{\Phi^{[+]}}\right)\sin\beta
\label{Eq:Wigcorr}
\end{eqnarray}
when we introduce the notation 
\begin{equation}
\hat{\mathcal W}(z,k)\equiv \hat D(z,k)\hat P\hat D^\dagger(z,k)
\label{Eq:Wop}
\end{equation}
for the shifted atomic parity according to Eq.~(\ref{Eq:atomop}).

The computation of the correlation function, Eq.~(\ref{Eq:Wigcorr}), is reduced to that of the different combinations $\bra{\Phi^{[\pm]}}\hat {\mathcal W}(z,k)\ket{\Phi^{[\pm]}}$. It is a straight--forward calculation \cite{roy:1977,ban:1996} to show that for an arbitrary state $\ket\Phi$ the operator $\hat{\mathcal W}(z,k)$, Eq.~(\ref{Eq:Wop}) leads to an expectation value of the form
$$
\bra\Phi\hat {\mathcal W}(z,k)\ket\Phi= \int\limits_{-\infty}^\infty dy \,e^{-2iky} \Phi^*(z-y)\Phi(z+y)
$$
which is in fact the Wigner function \cite{schl:2001} up to a normalization factor. A closer investigation for the partial wave packets $\Phi^{[\pm]}(z,t)$, Eq.~(\ref{Gl:Phirechts}), shows that the resulting Wigner functions are as well Gaussian distributions, now both in position-- and in momentum--space. The explicit expressions appearing in the correlation function, Eq.~(\ref{Eq:Wigcorr}), read 
\begin{eqnarray}
\label{Eq:Wplus}
\bra{\Phi^{[+]}}\hat {\mathcal W}(z,k)\ket{\Phi^{[+]}}&\equiv&{\mathcal W}^+(z,k) \\ \nonumber
&=&\exp\left[-2\alpha^2(k-k_0)^2-
\frac{1}{2\alpha^2}\left(z+z_0-\frac{\hbar k}{m}t +\frac{D}{2}\right)^2\right],
\end{eqnarray}
\begin{eqnarray}
\label{Eq:Wminus}
\bra{\Phi^{[-]}}\hat {\mathcal W}(z,k)\ket{\Phi^{[-]}}&\equiv&{\mathcal W}^-(z,k) \\ \nonumber 
&=&\exp\left[-2\alpha^2(k-k_0)^2-
\frac{1}{2\alpha^2}\left(z+z_0-\frac{\hbar k}{m}t -\frac{D}{2}\right)^2\right]
\end{eqnarray}
and for the interference term we find
\begin{eqnarray}
&&\frac{i}{2}\left(\bra{\Phi^{[+]}}\hat {\mathcal W}(z,k)\ket{\Phi^{[-]}}-\bra{\Phi^{[-]}}\hat {\mathcal W}(z,k)\ket{\Phi^{[+]}}\right)\equiv{\mathcal W}^{\rm int}(z,k)\nonumber \\ &=&\exp\left[-2\alpha^2(k-k_0)^2-
\frac{1}{2\alpha^2}\left(z+z_0-\frac{\hbar k}{m}t\right)^2\right]\sin[D(k-2k_0)].
\label{Eq:Wint}
\end{eqnarray}
These functions remain stationary around the momentum $k=k_0$ while in position space they move with velocity $(\hbar k)/m$ in positive $z$--direction and additionally get sheared. Moreover, each of the two Wigner functions ${\mathcal W}^+(z,k)$ and ${\mathcal W}^-(z,k)$, Eqs.~(\ref{Eq:Wplus}) and (\ref{Eq:Wminus}), is shifted by half of the distance $D$, Eq.~(\ref{Eq:distance}), originating from the interaction with the cavity field. In contrast to this the interference term ${\mathcal W}^{\rm int}(z,k)$, Eq.~(\ref{Eq:Wint}), remains centered but its momentum envelope oscillates with frequency $D$.

\begin{figure}[h]
\includegraphics[width=10cm]{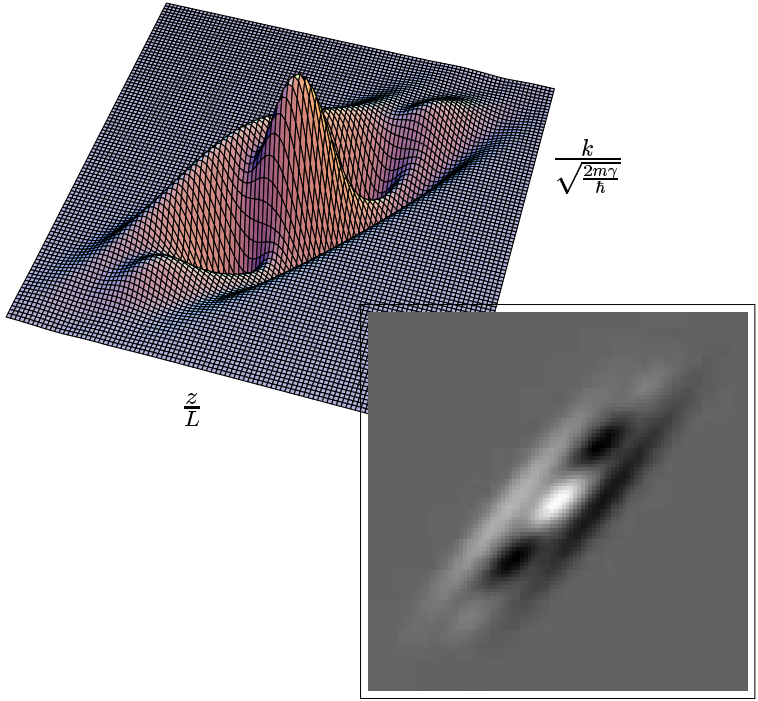}
\caption{Correlation function Eq.~(\ref{Eq:Wigcorr}) in phase--space. For a equal weight of the two Gaussian distributions ${\mathcal W}^{\pm}(z,k)$ and the interference term ${\mathcal W}^{\rm int}(z,k)$, i.e. for $\beta=\pi/4$, we get the typical picture of a Schr\"odinger--cat. For the generation of this plot we scaled all length with half the width $L$ of the resonator field, all wave numbers with $(2m\gamma/\hbar)^{1/2}$ and the time with $\gamma$, where $m$ is the mass of the atom and $\gamma$ the coupling--strength between atom and cavity field. The plot results from the parameter settings $(2m\gamma/\hbar)^{1/2}L=250$, $z_0/L=1.75$, $\alpha/L=0.03$, $k_0/(2m\gamma/\hbar)^{1/2}=2.31$. and $\gamma t=190$. The spatial separation between ${\mathcal W}^+(z,k)$ and ${\mathcal W}^-(z,k)$ is therefore given by $D/L=2(2m\gamma/\hbar)/k_0^2=0.375$.}
\label{fig:corrfkt}
\end{figure}

The correlation function $\langle\hat{\mathcal A}(z,k)\hat{\mathcal B}(\beta)\rangle$, Eq.~(\ref{Eq:Wigcorr}), therefore has a clear interpretation in phase space: Fig.~\ref{fig:corrfkt} shows the typical picture of a Schr\"odinger--cat in phase space which results from Eq.~(\ref{Eq:Wigcorr}) for equal--weighted trigonometric functions, i.e. for $\beta=\pi/4$. However, due to the special structure of the correlation function here the wave packet ${\mathcal W}^- (z,k)$ appears negative, whereas the wave packet ${\mathcal W}^+ (z,k)$ is always positive.

Additionally to the measurement parameters $z$, $k$ and $\beta$ the momentum $\hbar k_0$, the position $-z_0$ and the width $\alpha$ of the initial atomic wave packet, Eq.~(\ref{Eq:initgauss}), enter in the quantity $\langle\hat{\mathcal A}(z,k)\hat{\mathcal B}(\beta)\rangle$. Moreover, the measurement time $t>T$, the coupling strength $\gamma$ between atom and field and the length of the interaction zone $2L$ can be varied in a certain range. 

By scaling all lengths with half the width $L$ of the interaction zone $2L$, all wave numbers by $\sqrt{2m\gamma/\hbar}$ and additionally fixing the product $\sqrt{2m\gamma/\hbar}\cdot L$ of the two scaling--parameters, the correlation function, Eq.~(\ref{Eq:Wigcorr}), depends in all on eight parameters.

\subsection{Nonclassical behavior of the combined atom--field system}

The final step in order to identify the nonclassical regime of the CHSH--sum of correlations, Eq.~(\ref{Eq:qmCHSH}), is now to combine four of these correlation functions, Eq.~(\ref{Eq:Wigcorr}), that is
\begin{equation}
{\mathcal B}_{\rm QM}=\langle\hat{\mathcal A}(z,k)\hat{\mathcal B}(\beta)\rangle+\langle\hat{\mathcal A}(z,k)\hat{\mathcal B}(\beta')\rangle+\langle\hat{\mathcal A}(z',k')\hat{\mathcal B}(\beta)\rangle-\langle\hat{\mathcal A}(z',k')\hat{\mathcal B}(\beta')\rangle.
\label{Eq:violchsh}
\end{equation}
The CHSH--combination ${\mathcal B}_{\rm QM}$ depends on six real parameters: the positions $z$ and $z'$, the wave numbers $k$ and $k'$ and the phases $\beta$ and $\beta'$. These six parameters have to be varied in order to find nonclassical values ${\mathcal B}_{\rm QM}>2$, whereas the initial values of the wave packet, Eq.~(\ref{Eq:initgauss}), and the time $t>T$ of the measurement are fixed for all four correlations.

In our case the CHSH--combination, Eq.~(\ref{Eq:violchsh}) can be written as
\begin{equation}
{\mathcal B}_{\rm QM}= X_1\cdot\cos\beta+Y_1\cdot\sin\beta+X_2\cdot\cos\beta'+Y_2\cdot\sin\beta'
\label{helprel}
\end{equation}
where the $X_i$ and $Y_i$ depend on $z,z',k$ and $k'$.

This form allows us to maximize ${\mathcal B}_{\rm QM}$ with respect to the phases $\beta$ and $\beta'$. The full expression of ${\mathcal B}'_{\rm QM}\equiv \operatorname{max}\limits_{\beta,\beta'}{\mathcal B}_{\rm QM}$ can then be found by inserting the explicit expressions of the wave packets, Eqs.~(\ref{Eq:Wplus}) -- (\ref{Eq:Wint}). We find
\begin{eqnarray*}
{\mathcal B}'_{\rm QM}&=&\bigg{[}e^{-2d^2}\left(e^{-\kappa^2-x^2}\sinh[2xd]+e^{-\kappa'^2-x'^2}\sinh[2x'd]\right)^2 \\ &+&\left(e^{-\kappa^2-x^2}\sin[2d(\kappa-\kappa_0)]+e^{-\kappa'^2-x'^2}\sin[2d(\kappa'-\kappa_0)]\right)^2\bigg{]}^{\frac{1}{2}} \nonumber \\ &+&\bigg{[}e^{-2d^2}\left(e^{-\kappa^2-x^2}\sinh[2xd]-e^{-\kappa'^2-x'^2}\sinh[2x'd]\right)^2\\ &+&\left(e^{-\kappa^2-x^2}\sin[2d(\kappa-\kappa_0)]-e^{-\kappa'^2-x'^2}\sin[2d(\kappa'-\kappa_0)]\right)^2\bigg{]}^{\frac{1}{2}},
\end{eqnarray*}
where we have introduced the abbreviations $x=(z+z_0-\hbar k t/m)/(\sqrt{2}\alpha)$, $\kappa=\sqrt{2}\alpha(k-k_0)$, $d=D/(2\sqrt{2}\alpha)$ and $\kappa_0=\sqrt{2}\alpha k_0$ and the analogous expressions for the primed variables.

A numerical analysis of this expression shows the following result: The maximal value of ${\mathcal B}_{\rm QM}$ can be found for $\kappa=\kappa'=0$ and for $x=-x'=0.371$. In other words, the nonclassicality is maximal for wave numbers $k=k'=k_0$, i.e. for the initial momenta $\hbar k_0$, and for positions $x$ and $x'$ displaced symmetrically around zero. Inserting this values and optimizing the remaining parameters to $d=0.741$ and $\kappa_0=(2n+1)\cdot \pi/(4d)=(2n+1)\pi/2.954$, we find a maximal value ${\mathcal B}_{\rm QM}^{\rm max}= 2.324$. 

\begin{figure}[ht]
\centerline{\includegraphics[width=10cm]{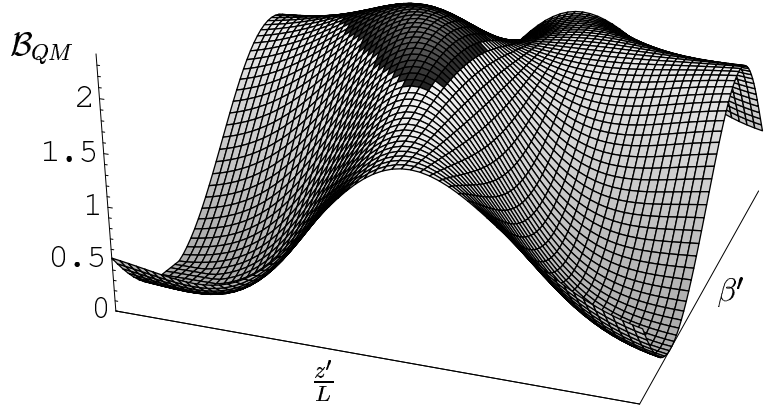}}
\caption{Nonclassical region of the CHSH--correlation sum, Eq.~(\ref{Eq:violchsh}). The CHSH--sum ${\mathcal B}_{\rm QM}$ is plotted over $z'/L$ and $\beta'$ while $z/L=0.24$ and $\beta=\pi/2$ are fixed. The parameters are $(2m\gamma/\hbar)^{1/2}L=6030.7$, $z_0/L=0.15$, $\alpha/L=0.067$, $k_0/(2m\gamma/\hbar)^{1/2}=2.668$ and $\gamma t=400$. The dyed region shows the nonclassical regime. For this parameters the maximal value is given by ${\mathcal B}_{\rm QM}^{\rm max}\approx 2.324$.}
\label{fig:viol}
\end{figure}

Fig.~\ref{fig:viol} shows the CHSH--sum as a function of the parameters $z'/L$ and $\beta'$ for $k=k'=k_0$, while we optimized the values of $z/L$ and $\beta$ numerically. The region with values ${\mathcal B}_{\rm QM}>2$ is dyed and shows for the chosen parameters a clear nonclassical regime of the CHSH--sum with a maximum of ${\mathcal B}_{\rm QM}\approx 2.324$.

The numerical optimization of ${\mathcal B}_{\rm QM}$ shows that the maximal value $2\sqrt 2$ allowed by quantum theory \cite{cir:1980} cannot be achieved. It seems that the special choice of the atomic observable $\hat{\mathcal A}(z,k)$, Eq.~(\ref{Eq:atomop}), leading to Gaussian--shaped correlation functions, is responsible for this fact: They smooth the CHSH--correlation, Eq.~(\ref{Eq:violchsh}), and therefore reduce the maximal possible value of $2\sqrt 2$.
\par


\section{Operational meaning of the observables}
\label{sec:measure}

In the previous section we have introduced the operators $\hat{\mathcal A}(z,k)$ and $\hat{\mathcal B}(\beta)$, Eqs.~(\ref{Eq:atomop}) and (\ref{Eq:fieldop}), for the atom and the field. A combination of four correlation functions has led us to the CHSH--correlation sum, Eq.~(\ref{Eq:violchsh}), which was shown to reach nonclassical values for a specific choice of the parameters of our system. This implies the possibility of a simultaneous measurement of the atomic operator $\hat{\mathcal A}$ and the field operator $\hat{\mathcal B}$. In the present section we shall shortly present the basic ideas how each of the occurring operators can be realized experimentally. 

We concentrate first on the atomic operator $\hat{\mathcal A}(z,k)$, Eq.~(\ref{Eq:atomop}), and recall that it consists of two parts, namely of the operator $\hat{\mathcal W}(z,k)$, Eq.~(\ref{Eq:Wop}), for the atomic center of mass motion and the operator $\hat\sigma$, Eq.~(\ref{Eq:intop}), for the internal atomic states. 

The suitable operator $\hat{\mathcal W}(z,k)$ for the atomic motion turned out to be the parity, displaced in phase--space by the amount $(z,k)$. Thereby, the phase--space point $(z,k)$ had to be chosen such that the CHSH--correlation sum moved to the nonclassical regime. In order to realize this operator experimentally we first recall how the expectation value $\bra{\Phi^{[\pm]}}\,\hat{\mathcal W}(0,0)\,\ket{\Phi^{[\pm]}}$ of the parity operator could be measured. Since consecutive energy eigenstates of a symmetric potential have alternating parity we could for example determine the parity of the atomic wave packet by bringing it into a symmetric potential and measuring its energy. However, we are not interested in the parity of the atomic wave packet at the origin but at the phase--space point $(z,k)$, determined by the displacement operator $\hat D(z,k)$, Eq.~(\ref{Eq:displop}). This operator acts on the atomic wave packets $\ket{\Phi^{[\pm]}}$. It is equivalent to an instantaneous displacement of the atom from the origin to the point $(z,k)$. We can as well imagine to displace the potential by the same amount in the opposite direction, which might be experimentally easier than displacing the atom.

The operator $\hat\sigma$, Eq.~(\ref{Eq:intop}), for the internal atomic states has the structure of a Pauli--spin matrix and describes a measurement of the atomic dipole. The experimental realization can be achieved by state--selective field ionization of the atoms \cite{ben:1994}. 

For the field measurement we take into account that we have at most one excitation in the cavity. Therefore, the operator $\hat{\mathcal B}(\beta)$, Eq.~(\ref{Eq:fieldop}), also has the structure of a Pauli--spin matrix, but with an additional phase. The experimental realization of this operator is equivalent to an electric--field--measurement of the cavity. Such measurements have been done by state--of--the--art homodyne techniques (see e.g. \cite{bra:1990}). In principle, the field is coupled out from the cavity and due to a beam splitter becomes entangled with a coherent state $\ket{|a|e^{i\beta}}$ of large amplitude. The differences of the intensities at the two outcomes of the homodyne interferometer finally determine the electric field of the cavity. The phase $\beta$ is thereby determined by the phase of the coherent state. 
\par


\section{Conclusions}
\label{sec:concl}

In this paper we have proposed a simple dynamical model and presented how its entangled continuous degrees of freedom could be used to exhibit the nonlocal aspect of quantum mechanics. We have investigated the operators suitable in this context and have given ideas how to define them operationally. In a further publication, we were able to show \cite{fre:2001} that for atoms in a linear potential, e.g. under the influence of gravity, this nonlocal feature remains still valid. However, there might exist other, maybe simpler operators which would indicate nonlocality as well. Moreover, it is not yet clear, if other criteria for nonclassical correlations \cite{sim:2000,dua:2000,rei:2000} can be applied to our model. The influence of decoherence effects \cite{Giul:1991} is another topic we have not yet taken into account in this context. 
\par


This work was financially supported by the Deutsche Forschungsgemeinschaft and by the European Community's Human Potential Programme under contract ``QUEST'' HPRN--CT--2000--00121. K.W.'s work was additionally partially supported by a KBN grant 2P03 B 02123.
\par



\end{document}